\documentclass[12pt]{article}
\usepackage{epsf}
\usepackage{amsmath,amssymb}

\usepackage[dvipdfmx]{graphicx}

\usepackage{cite}

\usepackage{comment}

\begin{document}

\newcommand{\lsim}{\stackrel{<}{_\sim}}
\newcommand{\gsim}{\stackrel{>}{_\sim}}

\newcommand{\rem}[1]{{$\spadesuit$\bf #1$\spadesuit$}}

\renewcommand{\theequation}{\thesection.\arabic{equation}}

\renewcommand{\thefootnote}{\fnsymbol{footnote}}
\setcounter{footnote}{0}

\begin{titlepage}

\def\thefootnote{\fnsymbol{footnote}}

\begin{center}

\hfill UT-16-16\\
\hfill April, 2016\\

\vskip .75in

{\Large \bf 

  Production and Decay of Di-photon Resonance \\
  at Future $e^+e^-$ Colliders \\

}

\vskip .5in

{\large 
  Hayato Ito and Takeo Moroi
}

\vskip 0.25in

\vskip 0.25in

{\em Department of Physics, University of Tokyo,
Tokyo 113-0033, Japan}

\end{center}
\vskip .5in

\begin{abstract}
  Motivated by the ATLAS and CMS announcements of the excesses of
  di-photon events, we discuss the production and decay processes of
  di-photon resonance at future $e^+e^-$ colliders.  We assume that
  the excess of the di-photon events at the LHC is explained by a
  scalar resonance decaying into a pair of photons.  In such a case,
  the scalar interacts with standard model gauge bosons and,
  consequently, the production of such a scalar is possible at the
  $e^+e^-$ colliders.  We study the production of the scalar resonance
  via the associated production with photon or $Z$, as well as via the
  vector-boson fusion, and calculate the cross sections of these
  processes.  We also study the backgrounds, and discuss the
  detectability of the signals of scalar production with various decay
  processes of the scalar resonance.  We also consider the case where
  the scalar resonance has an invisible decay mode, and study how the
  invisible decay can be observed at the $e^+e^-$ colliders.
\end{abstract}

\end{titlepage}

\renewcommand{\thepage}{\arabic{page}}
\setcounter{page}{1}
\renewcommand{\thefootnote}{\#\arabic{footnote}}
\setcounter{footnote}{0}

\section{Introduction}
\label{sec:intro}
\setcounter{equation}{0}

High energy $e^+ e^-$ linear colliders, like International $e^+e^-$
Linear Collider (ILC) \cite{Behnke:2013xla, Baer:2013cma,
  Adolphsen:2013jya, Adolphsen:2013kya, Behnke:2013lya} and
Compact Linear Collider (CLIC) \cite{Linssen:2012hp}, are attractive
candidates of energy frontier experiment in the future.  The future
$e^+e^-$ colliders will not only study the detail of standard-model
(SM) particles, particularly Higgs boson and top quark, but also
provide important information about new physics at the electroweak
scale (if it exists).  For the $e^+ e^-$ linear collider experiments
in the future, it is crucial to understand how and how well the
information about various new physics models can be obtained there.

Notably, in December of 2015, both the ATLAS and CMS collaborations
announced the excess in the di-photon invariant mass distribution
\cite{ATLAS-CONF-2015-081, CMS-PAS-EXO-15-004}.  The ATLAS observed
the excess with $3.6$-$\sigma$ ($2.0$-$\sigma$) local (global)
significance at $M_{\gamma\gamma}\simeq 750\ {\rm GeV}$ (with
$M_{\gamma\gamma}$ being the di-photon invariant mass) for a narrow
width case.  Furthermore, the CMS result shows an excess with the
local (global) significance of $2.6$-$\sigma$ ($1.2$-$\sigma$) at
$M_{\gamma\gamma}\simeq 760\ {\rm GeV}$.  These signals may indicate
the existence of a new scalar resonance with the mass of $\sim 750\
{\rm GeV}$, although more data is needed to confirm or exclude such a
possibility, (see, for example, \cite{Harigaya:2015ezk,
  Mambrini:2015wyu, Backovic:2015fnp, Angelescu:2015uiz,
  Nakai:2015ptz, Knapen:2015dap, Buttazzo:2015txu, Pilaftsis:2015ycr,
  Franceschini:2015kwy, DiChiara:2015vdm, Higaki:2015jag}).  If there
exists such a resonance, its properties should be studied in detail by
the future $e^+e^-$ collider experiments
\cite{Ito:2016zkz, Sonmez:2016xov, Djouadi:2016eyy, He:2016olo}.

Currently, the ILC is planned to be extendable up to $\sqrt{s}\sim 1\
{\rm TeV}$ (with $\sqrt{s}$ being the center-of-mass (CM) energy of
the collider).  In addition, the energy of CLIC can be as high as a
few TeV.  With such CM energies, the resonance with its mass of $\sim
750\ {\rm GeV}$ is kinematically reachable.  In particular, in some
class of models, the resonance can be produced in association with
neutral electroweak gauge bosons (i.e., $\gamma$ or $Z$) and via the
vector-boson fusion processes at the $e^+e^-$ colliders.  Once
produced, the properties of the resonance may be studied with a high
luminosity and a clean environment of the future $e^+e^-$ colliders.

If there exists a new resonance, it is important to understand how it
interacts with other fields.  As mentioned above, one attractive
explanation of the LHC di-photon excess is the existence of a scalar
resonance with its mass of $\sim 750\ {\rm GeV}$ coupled to the
standard-model (SM) gauge bosons.  In addition, as well as the
coupling to the gauge bosons, the scalar boson may also couple to
other fields.  For example, the scalar resonance may have a coupling
to an invisible new particle which may be dark matter of the universe
(see, for example, \cite{Mambrini:2015wyu, Backovic:2015fnp}).
Understanding of the properties of the scalar resonance will be a very
important issue if it exists.

The purpose of this paper is to consider the production and the decay
of the scalar boson (which we call $\Phi$), which is responsible for
the LHC di-photon excess, at the future $e^+e^-$ colliders.  We
calculate the production cross section of such a scalar resonance at
the $e^+e^-$ colliders.  We also estimate the number of backgrounds,
and discuss the detectability of each decay mode of $\Phi$.

The organization of this paper is as follows.  In Section
\ref{sec:model}, we summarize the model we consider.  In Section
\ref{sec:production}, we discuss the production processes of $\Phi$ at
the $e^+ e^-$ colliders.  In particular, we study the production of
$\Phi$ in association with $\gamma$ or $Z$, and also the production
via the vector-boson fusion processes.  In Section \ref{sec:sm}, we
consider the detectability of the $\Phi$ production signal in which
$\Phi$ decays into SM gauge bosons.  Detectability of the invisible
decay of $\Phi$ is discussed in Section \ref{sec:invisible}.  Section
\ref{sec:summary} is devoted for conclusions and discussion.

\section{Model}
\label{sec:model}
\setcounter{equation}{0}

Let us first summarize the interaction of the new scalar boson of our
interest.  In order to make our discussion concrete, we assume that
the scalar boson $\Phi$ is pseudo-scalar, and that it has the
following interaction
\begin{align}
  {\cal L}_{\rm eff} = 
  \frac{1}{2\Lambda_1} \Phi 
  \epsilon^{\mu\nu\rho\sigma} {\cal B}_{\mu\nu} {\cal B}_{\rho\sigma}
  + \frac{1}{2\Lambda_2} \Phi 
  \epsilon^{\mu\nu\rho\sigma} {\cal W}_{\mu\nu}^a {\cal W}_{\rho\sigma}^a
  + \frac{1}{2\Lambda_3} \Phi 
  \epsilon^{\mu\nu\rho\sigma} {\cal G}_{\mu\nu}^A {\cal G}_{\rho\sigma}^A,
  \label{L_pseudo}
\end{align}
where ${\cal B}_{\mu\nu}$, ${\cal W}_{\mu\nu}^a$, and ${\cal
  G}_{\mu\nu}^A$ are field strength tensors for $U(1)_Y$, $SU(2)_L$,
and $SU(3)_C$ gauge interactions, respectively, and the superscript
$a$ and $A$ are indices of the adjoint representations of $SU(2)_L$
and $SU(3)_C$, respectively.  (Even if $\Phi$ is a real scalar boson,
the following results are almost unchanged.)  The summations over the
repeated indices are implicit.  Because the interactions given in Eq.\
\eqref{L_pseudo} are non-renormalizable, it is expected that there
exists some dynamics which generates the interaction between $\Phi$
and SM gauge bosons.  We do not specify the dynamics behind the
effective Lagrangian, and use Eq.\ \eqref{L_pseudo} for our
study.\footnote
{If the energy of the collider becomes larger than the energy scale of
  the new physics responsible for the effective interaction, the
  calculation based on Eq.\ \eqref{L_pseudo} may be inaccurate.  We
  assume that it is not the case.  In particular, if the scale of
  generating ${\cal L}_{\rm eff}$ is close to
  $\sim\frac{1}{2}\sqrt{s}$, on the contrary, momentum-dependent
  corrections to the effective Lagrangian can become sizable.  Study
  of such an effect is interesting because it may reveal the dynamics
  behind the interaction of $\Phi$ with the SM gauge bosons.  Such an
  issue is, however, beyond the scope of our study, and we leave it
  for future study.}

With the above interaction terms, the partial decay rates of $\Phi$
into the gauge bosons are given by
\begin{align}
  \Gamma (\Phi\rightarrow gg) &= 
  \frac{2 m_\Phi^3}{\pi \Lambda_3^2},
  \\
  \Gamma (\Phi\rightarrow \gamma\gamma) &= 
  \frac{m_\Phi^3}{4 \pi \Lambda_{\gamma\gamma}^2},
  \\
  \Gamma (\Phi\rightarrow \gamma Z) &= 
  \frac{m_\Phi^3}{8 \pi \Lambda_{\gamma Z}^2}
  \left( 1 - \frac{m_Z^2}{m_\Phi^2} \right)^3,
  \\
  \Gamma (\Phi\rightarrow Z Z) &= 
  \frac{m_\Phi^3}{4 \pi \Lambda_{ZZ}^2}
  \left( 1 - \frac{4m_Z^2}{m_\Phi^2} \right)^{3/2},
  \\
  \Gamma (\Phi\rightarrow W^+ W^-) &= 
  \frac{m_\Phi^3}{2 \pi \Lambda_2^2}
  \left( 1 - \frac{4m_W^2}{m_\Phi^2} \right)^{3/2},
\end{align}
where $m_\Phi$, $m_Z$, and $m_W$ are the masses of $\Phi$, $Z$, and
$W^\pm$, respectively.  For the definitions of
$\Lambda_{\gamma\gamma}$, $\Lambda_{\gamma Z}$, and $\Lambda_{ZZ}$,
see Appendix.  As we have mentioned, we also consider the case where
$\Phi$ has an invisible decay mode.  In such a case, we treat the
invisible decay width $\Gamma (\Phi\rightarrow\chi\chi)$ as a free
parameter without specifying the interaction giving rise to such a
decay.  (Here and hereafter, the invisible particle is denoted as
$\chi$.)  One example is the interaction of the form of
$\Phi\chi\chi$, with which $\chi$ is regarded as a gauge singlet Weyl
fermion.  For the total decay width $\Gamma_\Phi$, because $\Phi$ may
decay into particles other than the SM gauge bosons or the invisible
particle, $\Gamma_\Phi$ is regarded as a free parameter.  In addition,
we assume that $\Gamma_\Phi\ll m_\Phi$ so that the narrow width
approximation is applicable.

In the following, we consider two possible production processes at the
LHC.  One is the gluon-gluon fusion process, for which the LHC cross
section (with the CM energy of $13\ {\rm TeV}$) is estimated as
\begin{align}
  \sigma_{{\rm LHC}}^{(gg)} (pp\rightarrow\Phi\rightarrow\gamma\gamma)
  \simeq 6.6\ {\rm fb} \times
  \frac{\Gamma (\Phi\rightarrow gg)}{\Gamma_\Phi}
  \times 
  \left[
    \frac{\Gamma (\Phi\rightarrow \gamma\gamma)}{1\ {\rm MeV}}
  \right],
  \label{sig_gg(PhiV)}
\end{align}
and the other is photon-photon fusion process \cite{Csaki:2015vek,
  Csaki:2016raa}, for which
\begin{align}
  \sigma_{{\rm LHC}}^{(\gamma\gamma)}
  (pp\rightarrow\Phi\rightarrow\gamma\gamma)
  \simeq 24\ {\rm fb} \times
  \left[
    \frac{\Gamma_\Phi}{100\ {\rm MeV}}
  \right]^{-1}
  \times
  \left[ 
    \frac{\Gamma (\Phi\rightarrow \gamma\gamma)}{100\ {\rm MeV}}
  \right]^2.
  \label{sig_gmmgmm(PhiV)}
\end{align}
One should note that, with the di-photon production cross section at
the LHC being fixed, a larger value of $\Gamma (\Phi\rightarrow
\gamma\gamma)$ is needed for the case of photon-photon fusion
dominance compared to the gluon-gluon fusion dominance.  This fact has
an important implication to the $e^+e^-$ colliders.

\section{Production of $\Phi$ at the $e^+e^-$ Colliders}
\label{sec:production}
\setcounter{equation}{0}

With the interaction given in Eq.\ \eqref{L_pseudo}, the $\Phi$
production may occur at the $e^+e^-$ colliders via several processes.
We first consider the production processes in association with
neutral electroweak gauge bosons:
\begin{itemize}
\item $e^+ e^- \rightarrow \Phi \gamma$,
\item $e^+ e^- \rightarrow \Phi Z$.
\end{itemize}
Feynman diagrams contributing to these processes are shown in Fig.\
\ref{fig:prodILC_sv}.  The analytic expressions of the cross sections
of these processes are given in Appendix.  In order to discuss the
production process of $\Phi$ at the $e^+e^-$ colliders in the light of
the LHC di-photon excess, it is convenient to parameterize the cross
sections of these processes by using the LHC cross sections.  Notably,
the ratio $\sigma (e^+ e^- \rightarrow \Phi
V)/\Gamma(\Phi\rightarrow\gamma\gamma)$ depends only on the ratio
$\Lambda_1/\Lambda_2$ (where $V=\gamma$ or $Z$).  Then, using Eq.\
\eqref{sig_gg(PhiV)}, one can see that the following relation holds:
\begin{align}
  \sigma (e^+ e^- \rightarrow \Phi V) Br(\Phi\rightarrow F)
  \simeq \bar{\sigma}^{(gg)}_{\Phi V}
  \times
  \frac{\Gamma (\Phi\rightarrow F)}{\Gamma (\Phi\rightarrow gg)}
  \times
  \left[
    \frac{\sigma_{{\rm LHC}}^{(gg)} (pp\rightarrow\Phi\rightarrow\gamma\gamma)}
    {10\ {\rm fb}}
  \right],
  \label{sigma_gg}
\end{align}
where $\bar{\sigma}_{gg}^{(V)}$ depends only on the ratio
$\Lambda_1/\Lambda_2$ (as far as $m_\Phi$ and $\sqrt{s}$ are
fixed). The above relation is useful when the LHC di-photon excess
originates from the gluon-gluon fusion.  In addition, using Eq.\
\eqref{sig_gmmgmm(PhiV)}, we define
$\bar{\sigma}^{(\gamma\gamma)}_{\Phi V}$, which is also a function of
$\Lambda_1/\Lambda_2$, as
\begin{align}
  \sigma (e^+ e^- \rightarrow \Phi V) Br(\Phi\rightarrow F)
  \simeq \bar{\sigma}^{(\gamma\gamma)}_{\Phi V}
  \times
  \frac{\Gamma (\Phi\rightarrow F)}{\Gamma (\Phi\rightarrow\gamma\gamma)}
  \times
  \left[
    \frac{\sigma_{{\rm LHC}}^{(\gamma\gamma)}
      (pp\rightarrow\Phi\rightarrow\gamma\gamma)}
    {10\ {\rm fb}}
  \right].
  \label{sigma_gmmgmm}
\end{align}
This expression can be used when the photon-photon fusion process is
important at the LHC.

\begin{figure}[t]
  \centering
  \includegraphics[height=0.2\textheight]{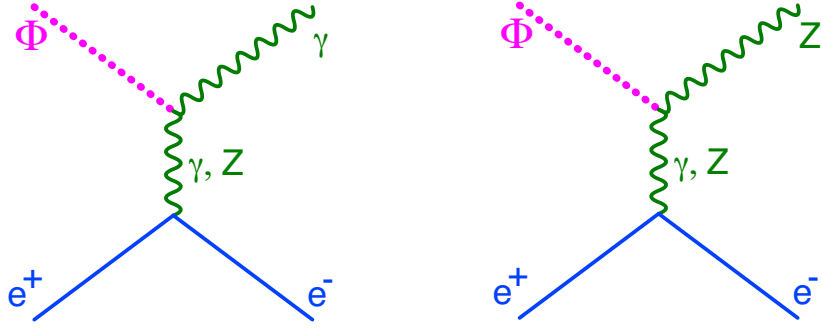}
  \caption{\small The Feynman diagrams for the processes $e^+ e^-
    \rightarrow \Phi \gamma$ and $\Phi Z$.}
  \label{fig:prodILC_sv}
\end{figure}

\begin{figure}
  \begin{center}
    \includegraphics[height=0.4\textheight]{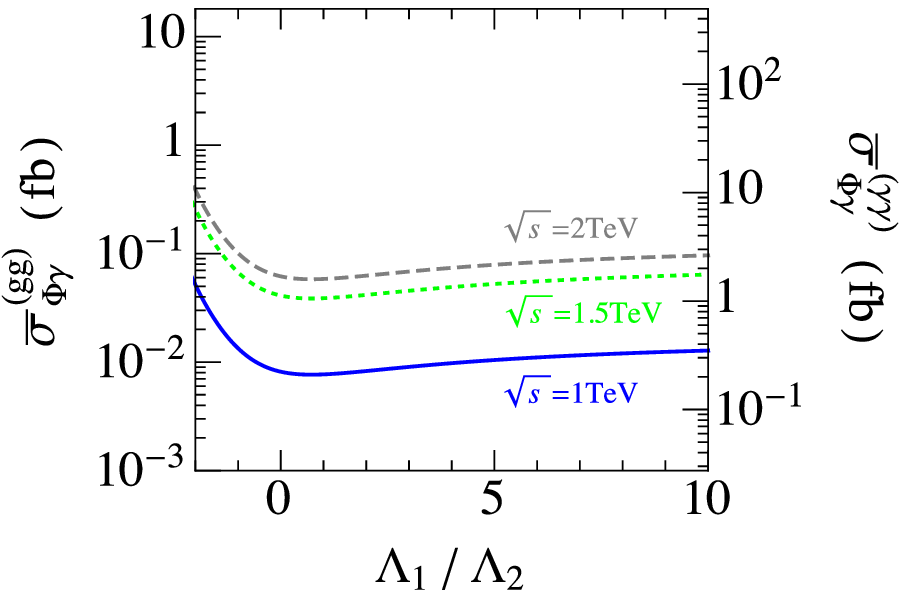}
    \caption{\small $\bar{\sigma}^{(gg)}_{\Phi \gamma}$ as a function
      of $\Lambda_1/\Lambda_2$, with $\sqrt{s}=1$, $1.5$, and $2\ {\rm
        TeV}$.  Here we take $P_{e-}=P_{e+}=0$.  The right-horizontal
      axis shows the value of $\bar{\sigma}^{(\gamma\gamma)}_{\Phi
        \gamma}$ using Eq.\ \eqref{gamma/g}.}
    \label{fig:sigma_agg}
    \includegraphics[height=0.4\textheight]{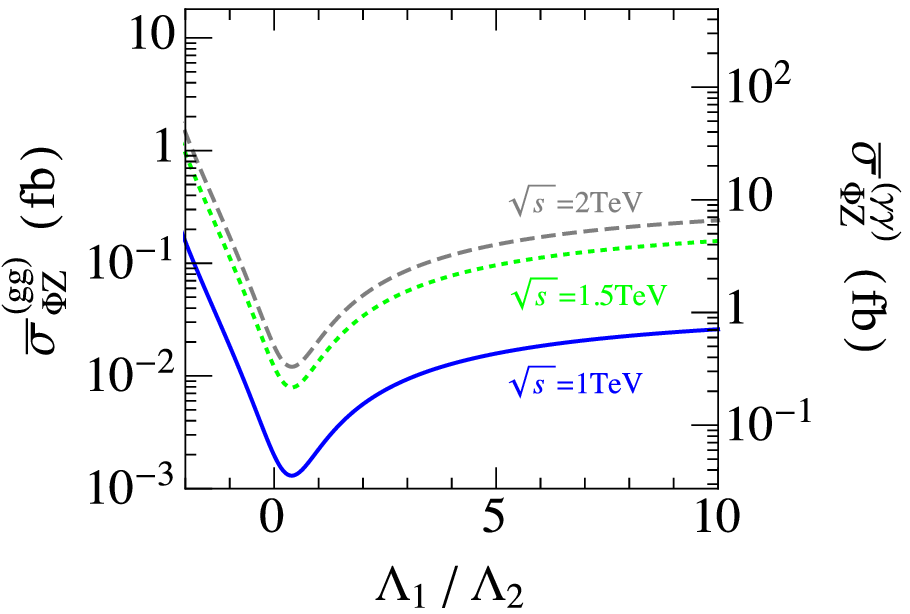}
    \caption{\small $\bar{\sigma}^{(gg)}_{\Phi Z}$ as a function of
      $\Lambda_1/\Lambda_2$, with $\sqrt{s}=1$, $1.5$, and $2\ {\rm
        TeV}$.  Here we take $P_{e-}=P_{e+}=0$.  The right-horizontal
      axis shows the value of $\bar{\sigma}^{(\gamma\gamma)}_{\Phi Z}$
      using Eq.\ \eqref{gamma/g}.}
    \label{fig:sigma_zgg}
  \end{center}
\end{figure}

In Figs.\ \ref{fig:sigma_agg} and \ref{fig:sigma_zgg}, we plot
$\bar{\sigma}^{(gg)}_{\Phi \gamma}$ and $\bar{\sigma}^{(gg)}_{\Phi Z}$
as functions of $\Lambda_1/\Lambda_2$, taking $\sqrt{s}=1$, $1.5$, and
$2\ {\rm TeV}$.  (Here and hereafter, we take $m_\Phi=750\ {\rm GeV}$
for our numerical calculations.)  Here, the electron and positron
beams are unpolarized, i.e., $P_{e-}=P_{e+}=0$ (with $P_{e^-}$ and
$P_{e^+}$ being the mean helicities of the initial-state electron and
positron, respectively).  Notice that the regions with too small or
too large $\Lambda_1/\Lambda_2$ are excluded by the $8\ {\rm TeV}$ run
of the LHC.  The most stringent bound comes from the negative searches
for the resonance which decays into $\gamma Z$; for example, taking
$\sigma (pp\rightarrow \Phi\rightarrow \gamma Z; 8\ {\rm TeV})< 11\
{\rm fb}$ \cite{Franceschini:2015kwy, Aad:2014fha} and the LHC cross
section of the di-photon signal events to be $10\ {\rm fb}$, only the
region with $-1\lesssim\Lambda_1/\Lambda_2\lesssim 6$ is allowed.  In
such a region, $\bar{\sigma}^{(gg)}_{\Phi \gamma}$ and
$\bar{\sigma}^{(gg)}_{\Phi Z}$ are both of $O(10^{-2}\ {\rm fb})$ or
smaller with $\sqrt{s}=1\ {\rm TeV}$.  With larger CM energy of $\sim
1.5-2\ {\rm TeV}$, the cross sections may be as large as $\sim 0.1\
{\rm fb}$.

\begin{figure}[t]
  \begin{center}
    \includegraphics[height=0.2\textheight]{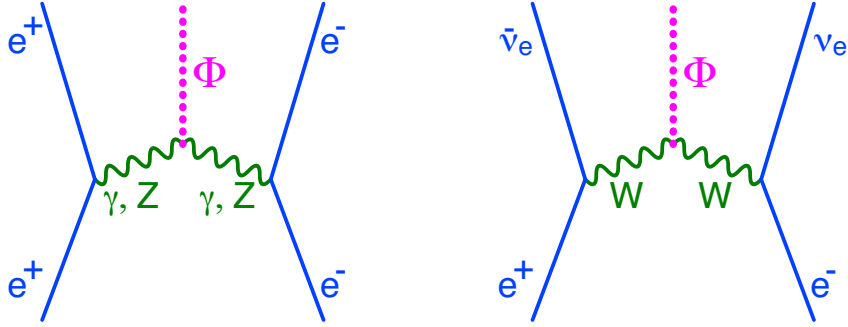}
    \caption{\small The vector-boson fusion diagrams contributing to the
      $\Phi$ productions.}
    \label{fig:prodILC_see}
    \end{center}
\end{figure}

Next, we consider the processes:
\begin{itemize}
\item $e^+ e^- \rightarrow \Phi e^+ e^-$,
\item $e^+ e^- \rightarrow \Phi \bar{\nu_e} \nu_e$,
\end{itemize}
to which the vector-boson fusion diagrams contribute (see Fig.\
\ref{fig:prodILC_see}).  For these processes, we define
\begin{align}
  \sigma (e^+ e^- \rightarrow \Phi \bar{l}l) Br(\Phi\rightarrow F)
  \simeq \bar{\sigma}^{(gg)}_{\Phi\bar{l}l}
  \times
  \frac{\Gamma (\Phi\rightarrow F)}{\Gamma (\Phi\rightarrow gg)}
  \times
  \left[
    \frac{\sigma_{{\rm LHC}}^{(gg)} (pp\rightarrow\Phi\rightarrow\gamma\gamma)}
    {10\ {\rm fb}}
  \right],
  \label{sigma_ll_gg}
\end{align}
and 
\begin{align}
  \sigma (e^+ e^- \rightarrow \Phi \bar{l}l) Br(\Phi\rightarrow F)
  \simeq \bar{\sigma}^{(\gamma\gamma)}_{\Phi\bar{l}l}
  \times
  \frac{\Gamma (\Phi\rightarrow F)}{\Gamma (\Phi\rightarrow\gamma\gamma)}
  \times
  \left[
    \frac{\sigma_{{\rm LHC}}^{(\gamma\gamma)}
      (pp\rightarrow\Phi\rightarrow\gamma\gamma)}
    {10\ {\rm fb}}
  \right],
  \label{sigma_ll_gmmgmm}
\end{align}
with $\bar{l}l=e^+e^-$ or $\bar{\nu}_e\nu_e$.  Then,
$\bar{\sigma}^{(gg)}_{\Phi\bar{l}l}$ and
$\bar{\sigma}^{(\gamma\gamma)}_{\Phi\bar{l}l}$ depend only on
$\Lambda_1/\Lambda_2$.

The cross section of the process $e^+ e^- \rightarrow \Phi e^+ e^-$ is
enhanced when the scattering angles of final-state $e^+$ and $e^-$ are
both small.  In such a case, the photon-photon fusion diagram shown in
Fig.\ \ref{fig:prodILC_see} is enhanced because the virtual photons
are almost on-shell so that the denominators of the photon propagators
become extremely small.  Then, the photon-photon fusion diagram
dominates over other diagrams which are less singular.  In order to
treat such an effect properly, we use the equivalent photon
approximation \cite{Budnev:1974de}; for the final state of $Fe^+e^-$,
we use
\begin{align}
  \sigma (e^+ e^- \rightarrow F e^+ e^-) \simeq
  \int dx dx'
  f_\gamma (x; \theta_{e^+}^{\rm (min)}, \theta_{e^+}^{\rm (max)})
  f_\gamma (x'; \theta_{e^-}^{\rm (min)}, \theta_{e^-}^{\rm (max)})
  \sigma (\gamma \gamma \rightarrow F; sxx'),
  \label{EqPhoton}
\end{align}
where $\theta_{e^\pm}^{\rm (min)}$ and $\theta_{e^\pm}^{\rm (max)}$
are minimal and maximal scattering angles of $e^\pm$, respectively,
and $\sigma (\gamma \gamma \rightarrow F; E_{\rm cm}^2)$ is the cross
section of the unpolarized photon-photon collision process $\gamma
\gamma \rightarrow F$ with the center-of-mass energy $E_{\rm cm}$.  In
addition, $f_\gamma$ is the distribution function of the photon; for
$0<x<1$, $f_\gamma$ is given by
\begin{align}
  f_\gamma (x,\theta^{\rm (min)}, \theta^{\rm (max)}) = 
  \frac{\alpha}{2\pi}
  \left[ \frac{1 + (1-x)^2}{x} \ln 
    \frac{|q^2|^{\rm (max)}}{|q^2|^{\rm (min)}}
    - 2 m_e^2 x 
    \frac{|q^2|^{\rm (max)}-|q^2|^{\rm (min)}}
    {|q^2|^{\rm (max)} |q^2|^{\rm (min)}}
  \right],
\end{align}
with $m_e$ being the electron mass, and 
\begin{align}
  |q^2|^{\rm (min,\, max)} \equiv
  \frac{1}{2} s (1-x)
  \left[
    1 - \cos\theta^{\rm (min,\, max)}
    + \frac{2 x^2}{(1-x)^2} \frac{m_e^2}{s}
  \right],
\end{align}
while, otherwise, $f_\gamma=0$.  For the process of our interest, we
obtain
\begin{align}
  \sigma (e^+ e^- \rightarrow \Phi e^+ e^-) \simeq &
  \frac{8\pi^2\Gamma (\Phi\rightarrow\gamma\gamma)}{s m_\Phi}
  \int \frac{dx}{x} 
  f_\gamma (x; \theta_{e^+}^{\rm (min)}, \theta_{e^+}^{\rm (max)})
  f_\gamma (m_\Phi^2 / sx; \theta_{e^-}^{\rm (min)}, \theta_{e^-}^{\rm (max)}),
\end{align}
where we used narrow width approximation.

The cross section of the process $e^+ e^- \rightarrow \Phi e^+ e^-$ is
logarithmically enhanced when $\theta_{e^\pm}^{\rm (min)}\ll 1$.  For
the study of the process $e^+e^-\rightarrow\Phi e^+e^-$, the energetic
$e^\pm$ in the forward direction may be used to eliminate the
backgrounds.  Since the ILC forward detectors are expected to cover up
to $O(1-10)$ mrad \cite{Behnke:2013lya}, we assume that energetic
$e^\pm$ with its energy $E_{e^\pm}$ larger than $50\ {\rm GeV}$ can be
identified with significant efficiency if the scattering angle
$\theta_{e^\pm}$ is larger than $10\ {\rm mrad}$.  We calculate the
cross section, requiring that $e^+$ and $e^-$ are emitted to the
forward directions.  We consider the following three different
requirements:
\begin{itemize}
\item Requirement 0: There is no $e^\pm$ with $\theta_{e^\pm}>100\
  {\rm mrad}$.  (In this case, the scattering angles of $e^\pm$ may be
  both so small that neither of $e^\pm$ are detected.)
\item Requirement 1: There is at least one $e^\pm$ with $E_{e^\pm}>50\
  {\rm GeV}$ and $\theta_{e^\pm}>10\ {\rm mrad}$.  In addition,
  $\theta_{e^\pm}<100\ {\rm mrad}$ is required for both $e^+$ and
  $e^-$.
\item Requirement 2: The energies and the scattering angles of both
  $e^+$ and $e^-$ satisfy $E_{e^\pm}>50\ {\rm GeV}$ and
  $10<\theta_{e^\pm}<100\ {\rm mrad}$.
\end{itemize}
The results are shown in Table \ref{table:sigma_eegg}.  (Notice that,
with the equivalent photon approximation, $\bar{\sigma}^{(gg)}_{\Phi
  e^+e^-}$ is independent of $\Lambda_1/\Lambda_2$.)  We have also
checked that, if we require $\theta_{e^\pm}>20\ {\rm mrad}$ for the
detection instead of $10\ {\rm mrad}$, $\bar{\sigma}^{(gg)}_{\Phi
  e^+e^-}$ decreases by $\sim 30\ \%$ and $\sim 50\ \%$ for the cases
of Requirement 2 and 3, respectively.

\begin{table}[t]
  \begin{center}
    \begin{tabular}{rrrr}
      \hline\hline
      $\sqrt{s}$ & Requirement 0 & Requirement 1 & Requirement 2
      \\
      \hline
      $1\ {\rm TeV}$
      & $0.044\ {\rm fb}$ & $0.015\ {\rm fb}$ & $0.002\ {\rm fb}$
      \\
      $1.5\ {\rm TeV}$
      & $0.18\ {\rm fb}$ & $0.064\ {\rm fb}$ & $0.007 \ {\rm fb}$
      \\
      $2\ {\rm TeV}$ & 
      $0.35\ {\rm fb}$ & $0.12 \ {\rm fb}$ & $0.012 \ {\rm fb}$
      \\
      \hline\hline
    \end{tabular}
    \caption{$\bar{\sigma}^{(gg)}_{\Phi e^+e^-}$ for $\sqrt{s}=1$,
      $1.5$, and $2\ {\rm TeV}$, adopting the Requirement 0, 1 or 2.
      $\bar{\sigma}^{(\gamma\gamma)}_{\Phi e^+e^-}$ can be obtained by
      using Eq.\ \eqref{gamma/g}.}
    \label{table:sigma_eegg}
  \end{center}
\end{table}

\begin{figure}[t]
  \centering
  \includegraphics[height=0.4\textheight]{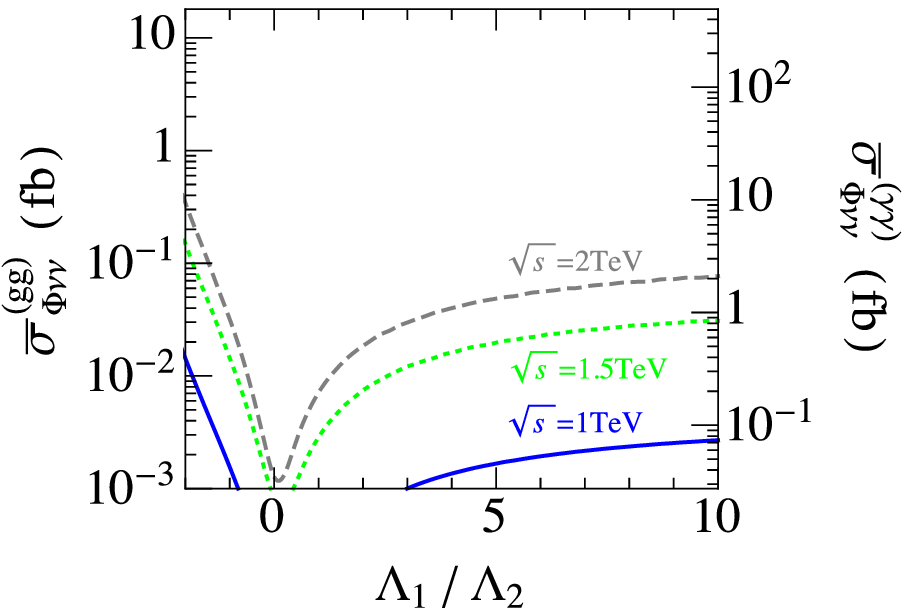}
  \caption{\small $\bar{\sigma}^{(gg)}_{\Phi\bar{\nu}_e\nu_e}$ as a
    function of $\Lambda_1/\Lambda_2$, with $\sqrt{s}=1$, $1.5$, and
    $2\ {\rm TeV}$.  Here we take $P_{e-}=P_{e+}=0$.  The
    right-horizontal axis shows the value of
    $\bar{\sigma}^{(\gamma\gamma)}_{\Phi\bar{\nu}_e\nu_e}$ using Eq.\
    \eqref{gamma/g}.}
  \label{fig:sigma_vvgg}
\end{figure}

For the process $e^+ e^- \rightarrow \Phi \bar{\nu_e} \nu_e$, the
$W$-boson fusion diagram contributes.  Even though $W$ is massive,
such a diagram is enhanced when the neutrinos are emitted to the
forward directions in the high energy limit.  Thus, the cross section
of such a process may potentially become larger than those for $e^+
e^- \rightarrow \Phi \bar{\nu}_\mu \nu_\mu$ and $\Phi \bar{\nu}_\tau
\nu_\tau$, which are given by $\sigma (e^+ e^- \rightarrow \Phi Z)$
multiplied by branching ratios of $Z$ into a neutrino pair.  We also
calculate $\bar{\sigma}^{(gg)}_{\Phi\bar{\nu}_e\nu_e}$, and the
results are shown in Fig.\ \ref{fig:sigma_vvgg}.  We found that
$\bar{\sigma}^{(gg)}_{\Phi\bar{\nu}_e\nu_e}$ is $O(10^{-3}\ {\rm fb})$
or smaller for $\sqrt{s}=1\ {\rm TeV}$.

So far, we have studied $\bar{\sigma}^{(gg)}_{\Phi X}$ (with
$X=\gamma$, $Z$, $e^+e^-$, or $\bar{\nu}_e\nu_e$).  For the
calculations of $\bar{\sigma}^{(\gamma\gamma)}_{\Phi X}$, we can use
the following proportionality:
\begin{align}
  \bar{\sigma}^{(\gamma\gamma)}_{\Phi X} \simeq
  27 \bar{\sigma}^{(gg)}_{\Phi X}.
  \label{gamma/g}
\end{align}
In particular, the values of $\bar{\sigma}^{(\gamma\gamma)}_{\Phi
  \gamma}$, $\bar{\sigma}^{(\gamma\gamma)}_{\Phi Z}$, and
$\bar{\sigma}^{(\gamma\gamma)}_{\Phi\bar{\nu}_e\nu_e}$ are also shown
in Figs.\ \ref{fig:sigma_agg}, \ref{fig:sigma_zgg}, and
\ref{fig:sigma_vvgg}, respectively.  When the LHC di-photon excess at
the LHC originates from the photon-photon fusion, the cross sections
at the $e^+e^-$ colliders become larger compared to the case of
gluon-gluon fusion at the LHC.  This is because, for the former case,
a larger value of $\Gamma (\Phi\rightarrow \gamma\gamma)$ needed (with
the di-photon cross section at the LHC being fixed), resulting in
stronger interaction of $\Phi$ with electroweak gauge bosons.

\section{Decay into SM Gauge Bosons}
\label{sec:sm}
\setcounter{equation}{0}

Now we are at the position to discuss the possibility of detecting the
$\Phi$ production at the future $e^+e^-$ colliders. In this section,
we consider the decay of $\Phi$ into SM gauge bosons.

\subsection{$e^+ e^-\rightarrow\Phi\gamma$ and $e^+ e^-\rightarrow\Phi Z$}

We first consider the $\Phi$ production in association with a SM gauge
boson, $e^+ e^-\rightarrow\Phi V$, followed by $\Phi\rightarrow V'_1
V'_2$, where $V=\gamma$ or $Z$, and $(V'_1,V'_2)=(g,g)$,
$(\gamma,\gamma)$, $(\gamma,Z)$, $(Z,Z)$, or $(W^+,W^-)$.  One
characteristic feature of such an event is the existence of a
monochromatic gauge boson.  With the process $e^+ e^-\rightarrow\Phi
V$, the energy of the gauge boson $V$ is given by
\begin{align}
  E_V^{\rm (sig)} = \frac{s - m_\Phi^2 + m_V^2}{2\sqrt{s}},
  \label{E_V}
\end{align}
where $m_V$ is the mass of $V$.  For the processes $e^+
e^-\rightarrow\Phi\gamma$ and $\Phi Z$, $E_\gamma^{\rm (sig)}=219\
{\rm GeV}$ and $E_Z^{\rm (sig)}=223\ {\rm GeV}$ ($859\ {\rm GeV}$ and
$861\ {\rm GeV}$) for $\sqrt{s}=1\ {\rm TeV}$ ($2\ {\rm TeV}$),
respectively.  The kinematical cut based on $E_V^{\rm (sig)}$ can be
used to reduce backgrounds.

To estimate the number of backgrounds, we calculate the SM cross
sections to produce $\gamma$ or $Z$ whose energy is close to $E_V^{\rm
  (sig)}$ in association with two other gauge bosons (or energetic
jets).  For simplicity, we do not consider the decay of weak bosons
nor the hadronization of partons.  Then, in studying the backgrounds
for the signal of $e^+e^-\rightarrow \Phi\gamma$, the following
kinematical requirement is imposed:
\begin{itemize}
\item There is one photon whose energy satisfies
  $|E_\gamma - E_\gamma^{\rm (sig)}|<0.02E_\gamma^{\rm (sig)}$, where
  $E_\gamma$ is the energy of photon.
\end{itemize}
For the backgrounds for $e^+e^-\rightarrow \Phi Z$, we require:
\begin{itemize}
\item There is one $Z$ whose energy satisfies
  $|E_Z - E_Z^{\rm (sig)}|<0.06E_Z^{\rm (sig)}$, where
  $E_Z$ is the energy of $Z$.
\end{itemize}
Notice that a very accurate measurement of the photon energy is
expected at the ILC \cite{Behnke:2013lya}; with the energy resolution
of the electromagnetic calorimeter of the SiD detector, for example,
$\delta E/E=0.17/\sqrt{E}\oplus 1\ \%$ for electrons or photons.
Furthermore, with the ILC detectors, the jet energy will be measured
with the accuracy of $3\ \%$ or better for jet energies above $100\
{\rm GeV}$.  Thus, we take $\sim 2$-$\sigma$ width of the detector
resolution, assuming that we use the hadronic decay mode of $Z$ for
the latter process.  In addition, we require that all the activities
satisfy
\begin{itemize}
\item $|\eta|<3$,
\end{itemize}
where $\eta$ is the pesudorapidity.

For the signal events in which $\Phi$ decays into a gluon pair, we
expect that the dominant source of the backgrounds is $e^+
e^-\rightarrow q\bar{q} V$, where $q$ denotes light quarks; we
calculate the SM cross sections of such processes with the cuts
mentioned above.  (In such a case, $(V'_1,V'_2)$ should be understood
as $(q,\bar{q})$.)  For other cases, we calculate the SM cross section
of the process $e^+ e^-\rightarrow V'_1 V'_2 V$.  The candidate of $V$
(i.e., the gauge boson produced in association with $\Phi$) is
expected to be identified by using Eq.\ \eqref{E_V}.  In addition, we
also assume that the final states with $(V'_1,V'_2)=(Z,Z)$ and
$(W^+,W^-)$ can be distinguished with hadronically decaying $Z$ and
$W^\pm$, using the invariant masses of the decay products of $V'_1$
and $V'_2$ (which we denote $m_{V'_1}$ and $m_{V'_2}$, respectively).
Assuming $3\ \%$ uncertainty for the measurement of jet energies, the
invariant masses of the $Z$- and $W^\pm$-systems are expected to be
determined with the accuracy of $\sim 4\ {\rm GeV}$, which is sizably
smaller than the mass difference of $Z$ and $W^\pm$.  In particular,
by studying $m_{V'_1}$ and $m_{V'_2}$ simultaneously, we expect that
the $(V'_1, V'_2)=(Z,Z)$ and $(W^+,W^-)$ final states are
distinguishable.\footnote
{The authors thank K. Fujii for pointing this out.}
The estimated numbers of backgrounds for the case of $\sqrt{s} = 1$
and $2$ TeV are summarized in Tables \ref{table:bg1TeV} and
\ref{table:bg2TeV}, respectively.

\begin{table}[t]
  \begin{center}
    \begin{tabular}{c|cccccccc}
      \hline\hline
      {Signal}
      & {$\gamma gg$}
      & {$Z gg$}
      & {$\gamma\gamma\gamma$}
      & {$\gamma\gamma Z$}
      & {$\gamma Z Z$}
      & {$Z Z Z$}
      & {$\gamma W^+ W^-$}
      & {$Z W^+ W^-$} \\
      \hline
      {$e^+e^-\rightarrow\Phi\gamma$} 
      & 840
      & $-$
      & 940
      & 670
      & 120
      & $-$
      & 1200
      & $-$ \\
      {$e^+e^-\rightarrow\Phi Z$} 
      & $-$
      & 900
      & $-$
      & 810
      & 550
      & 120
      & $-$
      & 3000 \\
      \hline\hline
    \end{tabular}
    \caption{\small The number of backgrounds for $e^+
      e^-\rightarrow\Phi\gamma$ and $\Phi Z$, with $\sqrt{s} = 1\ {\rm
        TeV}$ and $L=1\ {\rm ab}^{-1}$.  Here we take $P_{e^+} =
      P_{e^-} = 0$.}
    \label{table:bg1TeV}
\vspace{15pt}
    \begin{tabular}{c|cccccccc}
      \hline\hline
      {Signal}
      & {$\gamma gg$}
      & {$Z gg$}
      & {$\gamma\gamma\gamma$}
      & {$\gamma\gamma Z$}
      & {$\gamma Z Z$}
      & {$Z Z Z$}
      & {$\gamma W^+ W^-$}
      & {$Z W^+ W^-$} \\
      \hline
      {$e^+e^-\rightarrow\Phi\gamma$} 
      & 510
      & $-$
      & 1500
      & 890
      & 130
      & $-$
      & 1800
      & $-$ \\
      {$e^+e^-\rightarrow\Phi Z$} 
      & $-$
      & 700
      & $-$
      & 1800
      & 1100
      & 160
      & $-$
      & 3800 \\
      \hline\hline
    \end{tabular}
    \caption{\small Same as Table \ref{table:bg1TeV}, except for
      $\sqrt{s} = 2\ {\rm TeV}$.}
    \label{table:bg2TeV}
  \end{center}
\end{table}

To discuss the detectability of each mode, we define
\begin{align}
  S_{V V'_1 V'_2} / \sqrt{B_{V V'_1 V'_2}} \equiv
  \frac{L \sigma (e^+e^-\rightarrow \Phi V)
    Br (\Phi\rightarrow V'_1 V'_2) \epsilon}
  {\sqrt{B_{V V'_1 V'_2}}},
  \label{eq:significance}
\end{align}
where $L$ is the luminosity, $\epsilon$ is the efficiency due to the
rapidity cut, and $B_{V V'_1 V'_2}$ is the number of backgrounds for
the process $e^+e^-\rightarrow\Phi V$, followed by $\Phi\rightarrow
V'_1V'_2$.  Let us introduce
\begin{align}
  R_{\rm EW} \equiv 
  \frac{\Gamma (\Phi\rightarrow\gamma\gamma) + 
    \Gamma (\Phi\rightarrow\gamma Z) + \Gamma (\Phi\rightarrow Z Z) + 
    \Gamma (\Phi\rightarrow W^+ W^-)}{\Gamma (\Phi\rightarrow gg)},
\end{align}
which parametrizes the ratio of the gluon-gluon fusion and
photon-photon fusion contributions to the LHC di-photon events; with
sufficiently small (large) value of $R_{\rm EW}$, the LHC di-photon
excess is explained by the gluon-gluon (photon-photon) fusion process.
Then, the ratio $S_{V V'_1 V'_2} / \sqrt{B_{V V'_1 V'_2}}$ depends on
the luminosity $L$, $\Lambda_1/\Lambda_2$, $R_{\rm EW}$, and the LHC
di-photon cross section $\sigma_{{\rm
    LHC}}(pp\rightarrow\Phi\rightarrow\gamma\gamma)$, where
\begin{align}
  \sigma_{{\rm LHC}} (pp\rightarrow\Phi\rightarrow\gamma\gamma)
  = \sigma_{{\rm LHC}}^{(gg)}
  (pp\rightarrow\Phi\rightarrow\gamma\gamma)
  + \sigma_{{\rm LHC}}^{(\gamma\gamma)}
  (pp\rightarrow\Phi\rightarrow\gamma\gamma).
\end{align}
For our numerical analysis, we take $\sigma_{{\rm LHC}}
(pp\rightarrow\Phi\rightarrow\gamma\gamma)=10\ {\rm fb}$.  Using Eqs.\
\eqref{sig_gg(PhiV)} and \eqref{sig_gmmgmm(PhiV)}, the gluon-gluon and
photon-photon fusion contributions to the LHC di-photon cross section
become comparable when $\Gamma(\Phi\rightarrow\gamma\gamma)\simeq 27
\Gamma(\Phi\rightarrow gg)$.

First, let us consider the signal with $(V'_1,V'_2)=(g,g)$.  The ratio
$S_{Vgg} / \sqrt{B_{Vgg}}$ is maximized when the LHC di-photon excess
is explained by the gluon-gluon fusion process.  In such a case,
$S_{Vgg} / \sqrt{B_{Vgg}}$ depends only on the ratio $\Lambda_1 /
\Lambda_2$; with $\sqrt{s}=1\ {\rm TeV}$, it becomes larger than $5$
when $-4.9 < \Lambda_1 / \Lambda_2 < -2.5$ and $-8.2 < \Lambda_1 /
\Lambda_2 < -2.0$ for $V = \gamma$ and $V = Z$, respectively.  With
$\sigma_{{\rm LHC}} (pp\rightarrow\Phi\rightarrow\gamma\gamma)=10\
{\rm fb}$, however, such a value of $\Lambda_1 / \Lambda_2$ is already
excluded by the $8\ {\rm TeV}$ run of the LHC, as we have mentioned in
the previous section.  Thus, with the $\Phi$ production in association
with $\gamma$ or $Z$, we expect that the detection of $gg$ final state
is difficult.

\begin{figure}
  \centering
  \includegraphics[width=0.45\columnwidth]{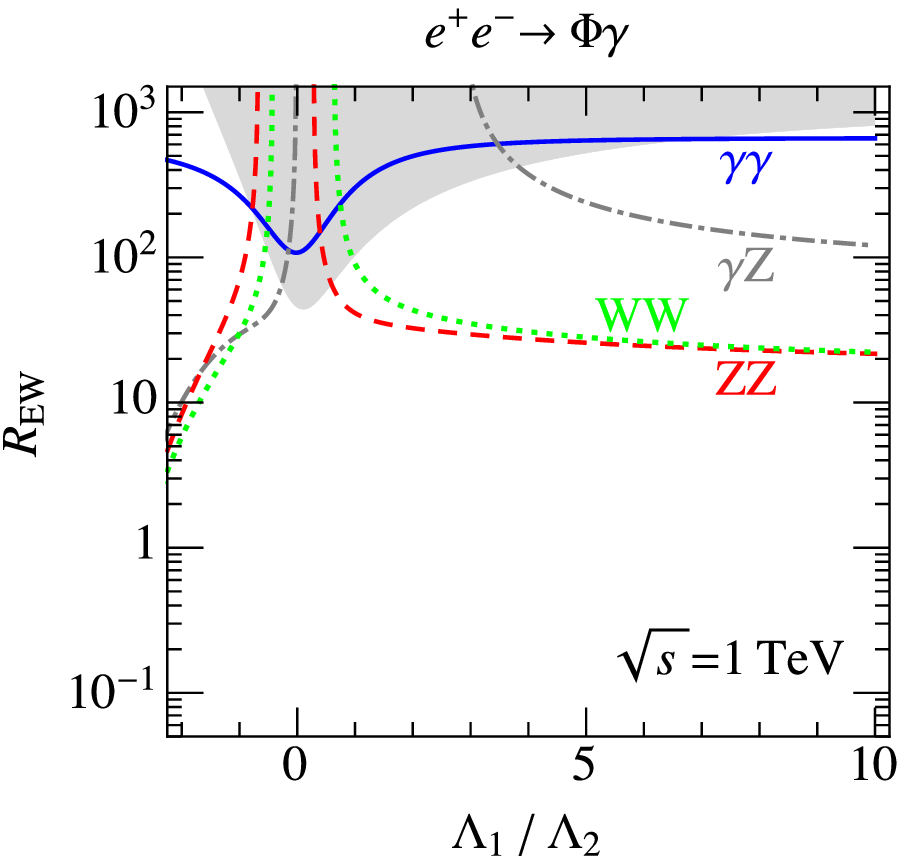}
  \hspace{0.05\columnwidth}
  \includegraphics[width=0.45\columnwidth]{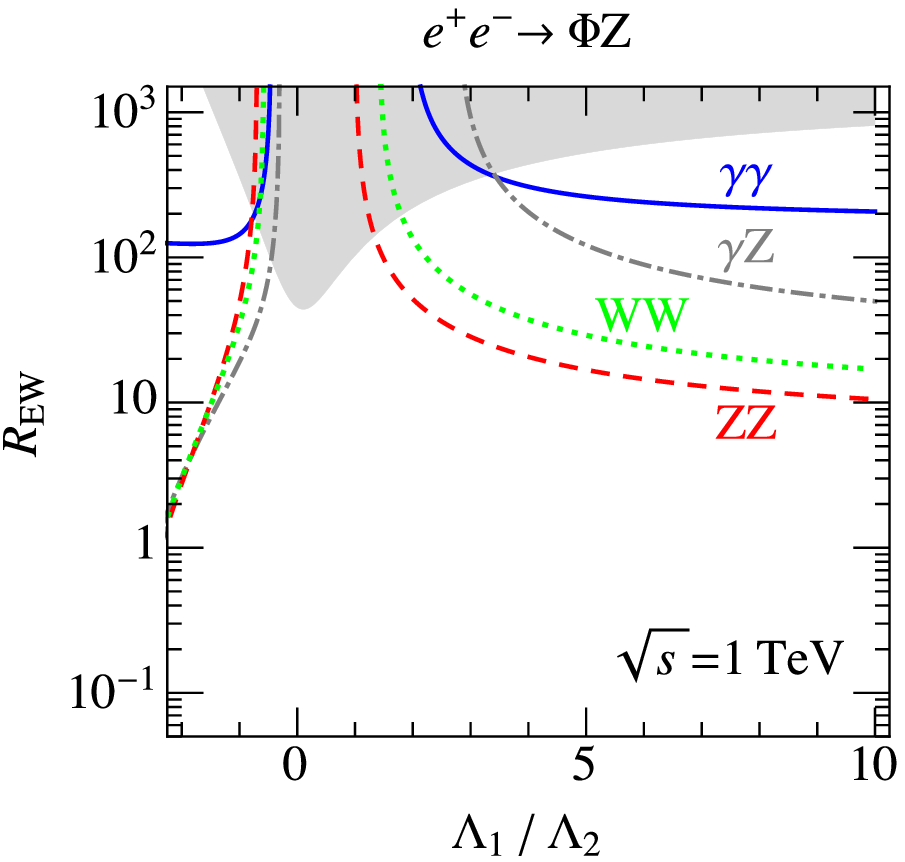}
  \caption{\small The minimal values of $R_{\rm EW}$ to realize $S_{V
      V'_1 V'_2} / \sqrt{B_{V V'_1 V'_2}} > 5$ for
    $V'_1V'_2=\gamma\gamma$ (blue solid), $\gamma Z$ (gray
    dot-dashed), $ZZ$ (red dashed), and $W^+W^-$ (green dotted) as
    function of the ratio $\Lambda_1 / \Lambda_2$.  Here we take
    $\sqrt{s}=1$ TeV, $L=1 \, {\rm ab}^{-1}$,
    $P_{e-}=P_{e+}=0$, 
    and $\sigma_{{\rm LHC}}
    (pp\rightarrow\Phi\rightarrow\gamma\gamma)=10\ {\rm fb}$.  In the
    shaded region, $\sigma_{{\rm LHC}}^{(\gamma\gamma)}
    (pp\rightarrow\Phi\rightarrow\gamma\gamma)$ is larger than
    $\sigma_{{\rm LHC}}^{(gg)}
    (pp\rightarrow\Phi\rightarrow\gamma\gamma)$.  }
  \label{fig:minR_1TeV}
  \vspace{7mm}
  \includegraphics[width=0.45\columnwidth]{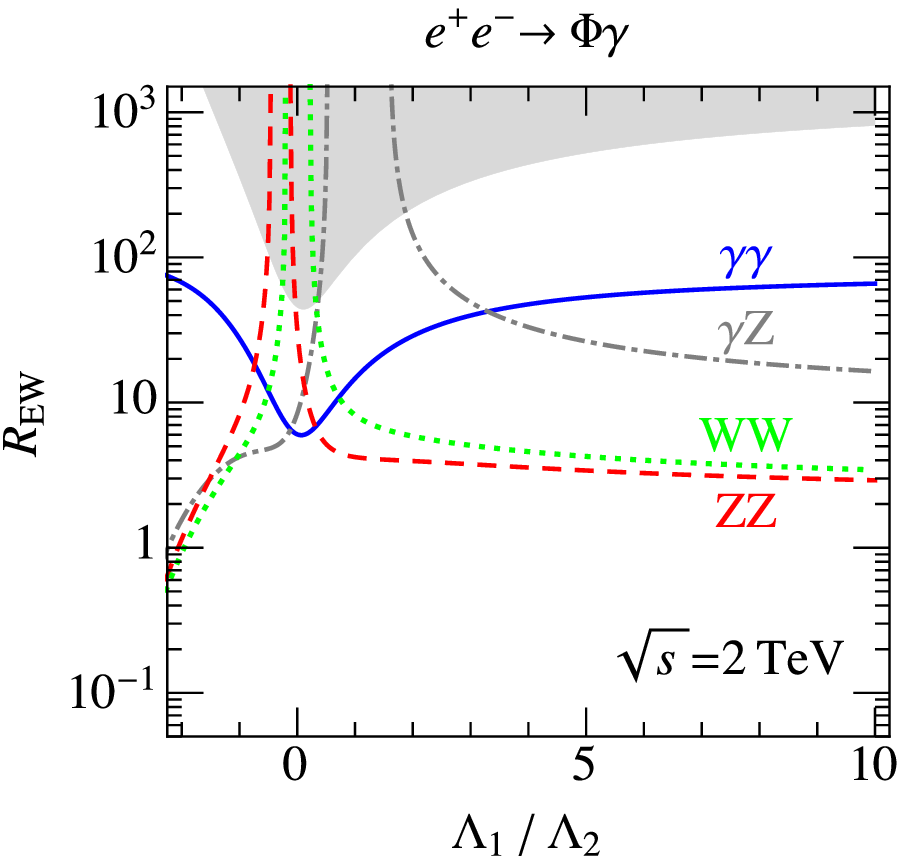}
  \hspace{0.05\columnwidth}
  \includegraphics[width=0.45\columnwidth]{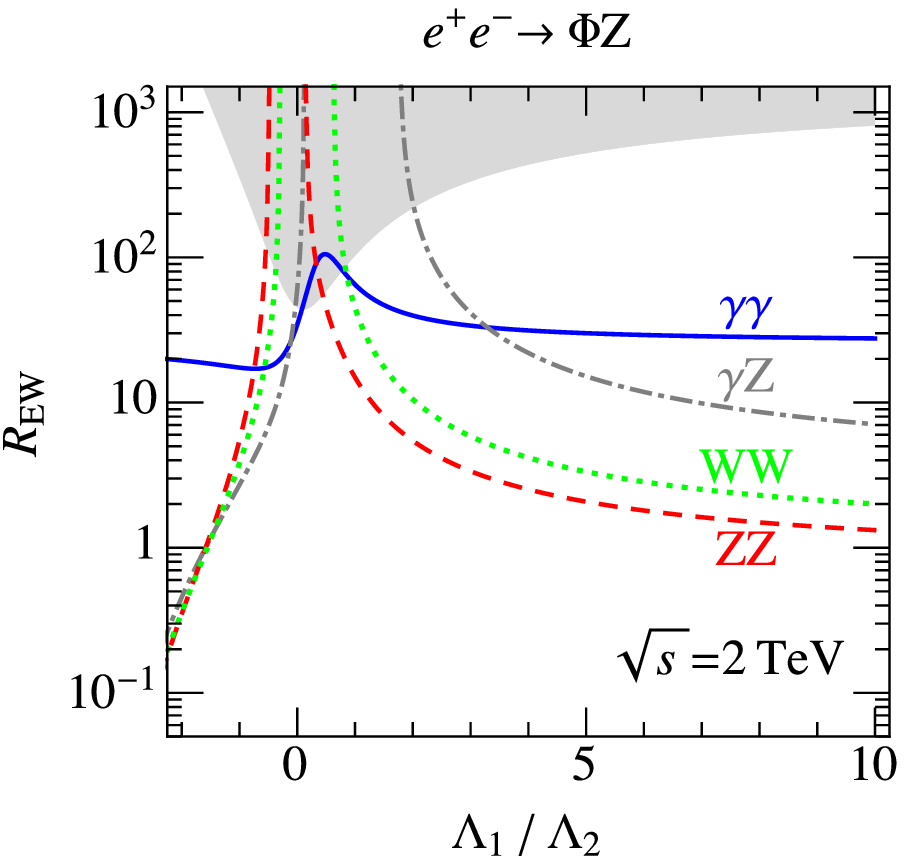}
  \caption{\small Same as Fig.\ \ref{fig:minR_1TeV}, except for
    $\sqrt{s}=2\ {\rm TeV}$.  }
  \label{fig:minR_2TeV}
\end{figure}

Next, we consider the other decay modes of $\Phi$, i.e.,
$\Phi\rightarrow V'_1 V'_2$ with $(V'_1,V'_2)=(\gamma,\gamma)$,
$(\gamma,Z)$, $(Z,Z)$, or $(W^+,W^-)$. As $R_{\rm EW}$ increases, the
cross sections of the $\Phi$ production with such decay processes
increase.  We estimate the minimal value of $R_{\rm EW}$ to see the
signals, requiring $S_{V V'_1 V'_2} / \sqrt{B_{V V'_1 V'_2}}>5$.  The
results are shown in Figs.\ \ref{fig:minR_1TeV} and
\ref{fig:minR_2TeV} for $\sqrt{s}=1$ and $2\ {\rm TeV}$, respectively.
In the figures, we shaded the region where $\sigma_{{\rm
    LHC}}^{(\gamma\gamma)} (pp\rightarrow\Phi\rightarrow\gamma\gamma)$
becomes larger than $\sigma_{{\rm LHC}}^{(gg)}
(pp\rightarrow\Phi\rightarrow\gamma\gamma)$; thus, in the shaded
region, the LHC di-photon excess is explained by the photon-photon
fusion process.  We can see that the detection and the study of $\Phi$
may be possible at the $e^+e^-$ colliders with large enough $R_{\rm
  EW}$.  In particular, if the LHC di-photon excess is due to the
photon-photon fusion, such a study seem possible even with
$\sqrt{s}=1\ {\rm TeV}$ and $L=1\ {\rm ab}^{-1}$.

\subsection{$e^+ e^-\rightarrow e^+ e^- \Phi$}

Next, we consider the process $e^+ e^-\rightarrow e^+ e^- \Phi$.  The
important feature of such a process is the existence of $e^\pm$ which
are (almost) parallel to the beam direction.  Detection of such
$e^\pm$ may help to reduce backgrounds.

We consider the case where at least one of the final-state $e^\pm$ is
detectable.  Then, as the signal, we require:
\begin{itemize}
\item Requirement 1 in the previous section: At least one $e^\pm$ with
  $E_{e^\pm}>50\ {\rm GeV}$ and $\theta_{e^\pm}>10\ {\rm mrad}$.  In
  addition, $\theta_{e^\pm}<100\ {\rm mrad}$ for both $e^+$ and $e^-$.
\item Candidates of $V'_1$ and $V'_2$, which are the gauge bosons
  produced by the decay of $\Phi$.  (Thus,
  $(V'_1,V'_2)=(\gamma,\gamma)$, $(g,g)$, $(\gamma,Z)$, $(Z,Z)$, or
  $(W^+,W^-)$.)
\end{itemize}
In the following, we estimate the number of SM backgrounds for this
type of events.  In order to reduce the backgrounds, we impose
kinematical selections based on the invariant mass of the $V'_1 V'_2$
system (which is denoted as $m_{V'_1 V'_2}$) and the pseudorapidities of
$V'_1$ and $V'_2$:
\begin{itemize}
\item $|m_{V'_1 V'_2} - m_{\Phi}| < 0.02 m_{\Phi}$
  for $(V'_1,V'_2)=(\gamma,\gamma)$, and 
  $|m_{V'_1 V'_2} - m_{\Phi}| < 0.06 m_{\Phi}$ otherwise.
\item $|\eta| < 1.47$ for $V'_1$ and $V'_2$.
\end{itemize}
Notice that, for the signal events with $\Phi\rightarrow gg$, we
expect that the dominant background is the process $e^+e^-\rightarrow
e^+e^-q\bar{q}$.  Thus, we also study the cross section of such a
process.

Now we estimate the number of SM backgrounds.  If there exists the
process $\gamma\gamma\rightarrow V'_1V'_2$, the cross section of
$e^+e^-\rightarrow e^+e^-V'_1V'_2$ is logarithmically enhanced when
the final-state $e^+$ and $e^-$ are both emitted to the forward
directions.  This is because diagrams containing $n$ nearly on-shell
photon propagators result in the cross section approximately
proportional to $\ln^n (|q^2|^{\rm (max)}/|q^2|^{\rm (min)})$, as we
discussed in the previous section (see Eq.\ \eqref{EqPhoton}).

We first consider the cases where the final states are $q\bar{q}$ or
$W^+W^-$.  In these cases, there exist tree-level processes
$\gamma\gamma\rightarrow q\bar{q}$ and $W^+W^-$.  Therefore, for the
backgrounds of these signal processes, diagrams with two nearly
on-shell photon propagators are expected to be the most important.  By
using the equivalent photon approximation, we estimate the SM cross
sections of the processes $e^+e^-\rightarrow e^+e^- q\bar{q}$ and
$e^+e^- W^+W^+$, imposing the above-mentioned kinematical selections.
The expected numbers of backgrounds for $L=1 \,{\rm ab}^{-1}$ are $47$
and $530$ for $(V'_1,V'_2)=(g,g)$ and $(W^+,W^-)$, respectively,
taking $\sqrt{s}=1$ TeV.  For $\sqrt{s}=2$ TeV, they are $250$ and
$2500$, respectively.  Since signal events with these final states
suffer from large numbers of backgrounds compared to the other final
states, as we will see below, we will not discuss further the
detectability of these signal events.

We now consider the signals with $\Phi\rightarrow V'_1V'_2$ with
$(V'_1,V'_2)=(\gamma,\gamma)$, $(\gamma, Z)$, and $(Z,Z)$.  For these
final states, the processes $\gamma\gamma\rightarrow V'_1V'_2$ occur
at the one-loop level, and are loop suppressed.  Assuming that the
one-loop processes with maximal logarithmic enhancements dominate the
backgrounds, we estimate the number of backgrounds using the
equivalent photon approximation; the cross sections of the processes
$\gamma\gamma\rightarrow \gamma\gamma$, $\gamma Z$, and $ZZ$ can be
found in \cite{Gounaris:1999gh, Gounaris:1999ux, Gounaris:1999hb}.
Then, the numbers of backgrounds are estimated to be 0.2, 3.1, and 5.1
(1.0, 17, and 28) for $(V'_1,V'_2)=(\gamma,\gamma)$, $(\gamma, Z)$,
and $(Z,Z)$, respectively, taking $L=1 \,{\rm ab}^{-1}$ and
$\sqrt{s}=1\ {\rm TeV}$ ($2\ {\rm TeV}$).\footnote
{In the tree-level diagrams contributing to the backgrounds of these
  processes, the number of nearly on-shell photon propagator is at
  most one.  However, the tree level contributions are potentially
  important because there is no loop suppression.  We have also
  studied the tree-level contributions by using {\tt
    MadGraph5\_aMC@NLO\,v2} \cite{Alwall:2014hca} and {\tt
    MadAnalysis} \cite{Conte:2012fm}.  For $10<\theta_{e^\pm}<100\
  {\rm mrad}$ for both $e^+$ and $e^-$, we directly calculated the
  cross section of the process $e^+e^-\rightarrow e^+e^- V'_1V'_2$.
  For $\theta_{e^+}<10\ {\rm mrad}$ and $10<\theta_{e^-}<100\ {\rm
    mrad}$, (or for $\theta_{e^-}<10\ {\rm mrad}$ and
  $10<\theta_{e^+}<100\ {\rm mrad}$), we adopted the equivalent photon
  approximation for the virtual photon emitted by $e^+$ (or $e^-$) and
  estimated the cross section.  For both regions of the phase space,
  we found that the number of backgrounds with $L=1\ {\rm ab}$ are much
  smaller than $1$.  Thus, we neglect the tree-level contributions.}

\begin{figure}
  \centering
  \includegraphics[width=0.45\columnwidth]{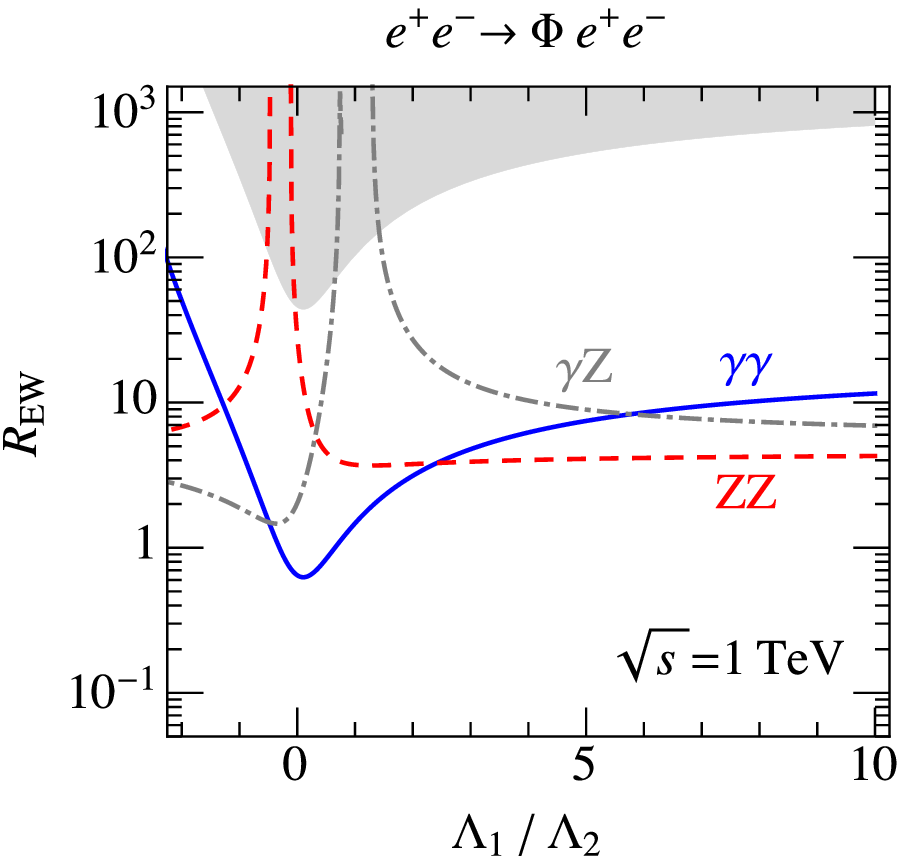}
  \hspace{0.05\columnwidth}
  \includegraphics[width=0.45\columnwidth]{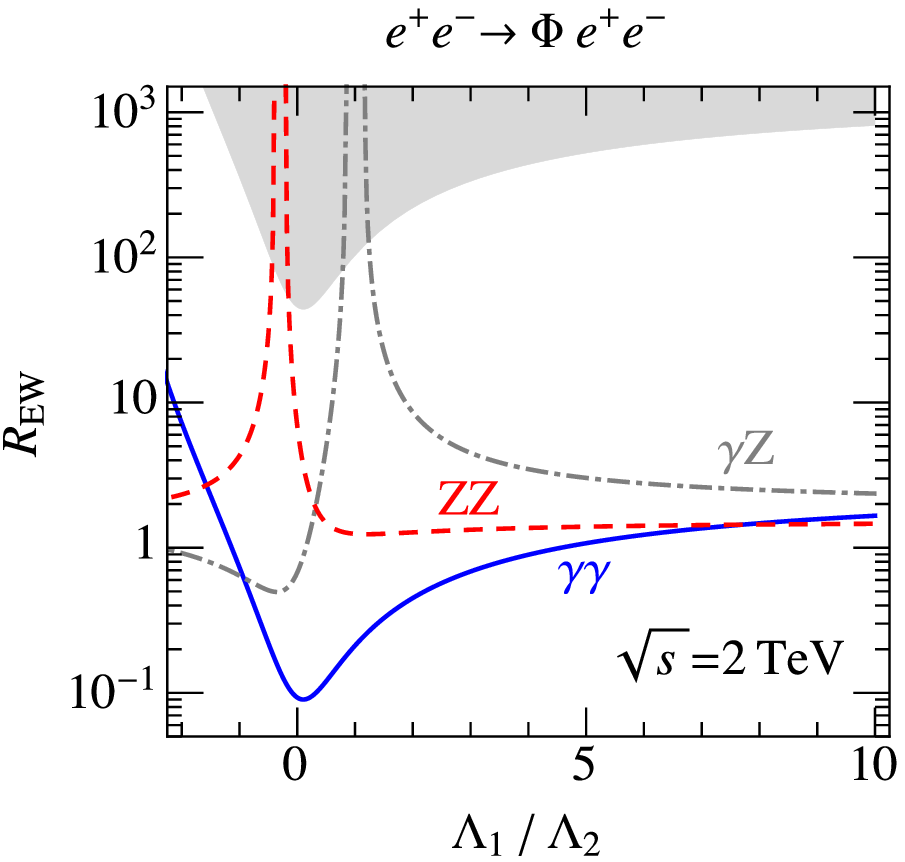}
  \caption{\small The minimal values of $R_{\rm EW}$ to realize $S_{V
      V'_1 V'_2} / \sqrt{B_{V V'_1 V'_2}} > 5$ for
    $V'_1V'_2=\gamma\gamma$ (blue solid), $\gamma Z$ (gray
    dot-dashed), and $ZZ$ (red dashed) as functions of the ratio
    $\Lambda_1 / \Lambda_2$.  In the shaded region, $\sigma_{{\rm
        LHC}}^{(\gamma\gamma)}
    (pp\rightarrow\Phi\rightarrow\gamma\gamma)$ is larger than
    $\sigma_{{\rm LHC}}^{(gg)}
    (pp\rightarrow\Phi\rightarrow\gamma\gamma)$.  Here we take
    $\sqrt{s}=1$ TeV (left) and $2$ TeV (right), $L=1 \, {\rm
      ab}^{-1}$, and $P_{e-}=P_{e+}=0$.  }
  \label{fig:minR_eeVV}
\end{figure}

With these background estimations, we calculate the minimal value of
$R_{\rm EW}$ which realizes $S_{eeV'_1V'_2} / \sqrt{B_{eeV'_1V'_2}}>
5$ for $L=1 \,{\rm ab}^{-1}$, where
\begin{align}
  S_{eeV'_1V'_2} \equiv L \sigma (e^+e^-\rightarrow e^+ e^- \Phi)
  Br (\Phi\rightarrow V'_1 V'_2) \epsilon,
\end{align}
while $B_{eeV'_1V'_2}$ is the number of backgrounds.  If
$B_{eeV'_1V'_2}$ is less than 1, we require $S_{eeV'_1V'_2} > 5$
instead.  In Figs.\ \ref{fig:minR_eeVV} we show the minimal values of
$R_{\rm EW}$ as functions of the ratio $\Lambda_1 / \Lambda_2$.
Comparing with Figs.\ \ref{fig:minR_1TeV} and \ref{fig:minR_2TeV}, we
can see that the process $e^+ e^-\rightarrow e^+ e^- \Phi$ is easier
to detect than $e^+ e^-\rightarrow \Phi V$ if $\Phi$ decays into a
pair of neutral electroweak gauge bosons.

\section{Invisible Decay}
\label{sec:invisible}
\setcounter{equation}{0}

In the previous section, we have considered the decay of $\Phi$ into
SM gauge bosons.  Notably, $\Phi$ may also couple to a new particle
which is not in the particle content of the SM.  In this section, we
consider the case where $\Phi$ couples to a particle which does not
have a direct coupling to SM particles.  In such a case, $\Phi$ has an
invisible decay mode; study of such a decay mode is an important step
to understand the property of $\Phi$.

One possibility of detecting the invisible decay of $\Phi$ at the
$e^+e^-$ colliders is to use the production process $e^+ e^-
\rightarrow \Phi \gamma$ and $\Phi Z$, followed by the invisible decay
of $\Phi$.  In such processes, we observe energetic $\gamma$ or the
decay products of $Z$ accompanied by large missing momentum.  In the
signal event, the energy of the SM gauge boson is $E_V^{\rm (sig)}$
given in Eq.\ \eqref{E_V}.  Thus, by selecting events with a single
$\gamma$ or $Z$ candidate whose energy is close enough to $E_V^{\rm
  (sig)}$, we may be able to eliminate backgrounds to observe the
invisible decay of $\Phi$.

For such a study, we estimate the number of backgrounds, applying cuts
on the energy of SM gauge bosons.  For the signal process $e^+ e^-
\rightarrow \Phi \gamma$ with $\Phi\rightarrow\chi\chi$, we expect a
single photon final state.  As possible sources of backgrounds, we
consider the following processes:
\begin{itemize}
\item[(i)] $e^+ e^-\rightarrow\gamma\bar{\nu}_l\nu_l$,
\item[(ii)] $e^+e^-\rightarrow\gamma \bar{\nu}_l\nu_l \bar{\nu}_{l'}\nu_{l'}$,
\end{itemize}
with $l,\, l' = e,\, \mu,$ and $\tau$; we calculate the cross sections
of these processes requiring that the energy of $\gamma$ is in the
range of $|E_\gamma - E_\gamma^{\rm (sig)}|<0.02E_\gamma^{\rm (sig)}$,
and that $|\eta|<1$ for the final-state photon.  For the signal
process $e^+ e^- \rightarrow \Phi Z$ with $\Phi\rightarrow\chi\chi$,
monochromatic $Z$-boson should be tagged to identify the signal.
Here, we use the hadronic decay of $Z$ because the hadronic branching
ratio is larger than leptonic one.  Then, we estimate the number of
backgrounds by calculating the cross sections of the following
processes:
\begin{itemize}
\item[(iii)] $e^+ e^-\rightarrow Z \bar{\nu}_l\nu_l$,
\item[(iv)] $e^+ e^-\rightarrow Z \bar{\nu}_l\nu_l \bar{\nu}_{l'}\nu_{l'}$,
\end{itemize}
where we require that $|E_{Z} - E_Z^{\rm (sig)}|<0.06E_\gamma^{\rm
  (sig)}$, and that $|\eta|<1$ for the $Z$-boson.  We use {\tt
  MadGraph5\_aMC@NLO\,v2} \cite{Alwall:2014hca} and {\tt MadAnalysis}
\cite{Conte:2012fm} to calculate the cross sections of the background
processes.  The numbers of backgrounds of these processes with the
luminosity of $1\ {\rm ab}^{-1}$ are summarized in Table
\ref{table:bg_invis}, taking $(P_{e^+}, P_{e^-}) =(-0.3, +0.8)$ and
$(+0.3, -0.8)$.  (We checked that one of these combinations of the
helicities gives the best detectability, as far as $|P_{e^-}|=0.8$ and
$|P_{e^+}|=0.3$.)

\begin{table}[t]
  \begin{center}
    \begin{tabular}{c|cccc}
      \hline\hline
      {}
      & {$\gamma + 2 \nu$}
      & {$\gamma + 4 \nu$}
      & {$Z + 2 \nu$}
      & {$Z + 4 \nu$}\\
      \hline
      $\sqrt{s}=1\ {\rm TeV}$
      & $560$ ($9100$)
      & $3.6$ ($37$)
      & $4800$ ($80000$)
      & $22$ ($330$)
      \\
      $\sqrt{s}=2\ {\rm TeV}$
      & $78$ ($1100$)
      & $1.9$ ($9.3$)
      & $390$ ($6000$)
      & $5.1$ ($35$)\\
      \hline\hline
    \end{tabular}
    \caption{\small The number of backgrounds for $e^+
      e^-\rightarrow\Phi\gamma$ and $\Phi Z$ followed by
      $\Phi\rightarrow \chi\chi$, with $\sqrt{s} = 1$ and 2 TeV, and
      $L=1\ {\rm ab}^{-1}$.  Here we take $P_{e^+} =-0.3\, (+0.3)$ and
      $ P_{e^-} = +0.8\, (-0.8)$ for the left (right) of each column.}
    \label{table:bg_invis}
    \end{center}
\end{table}

\begin{figure}
  \centering
  \includegraphics[height=0.4\textheight]{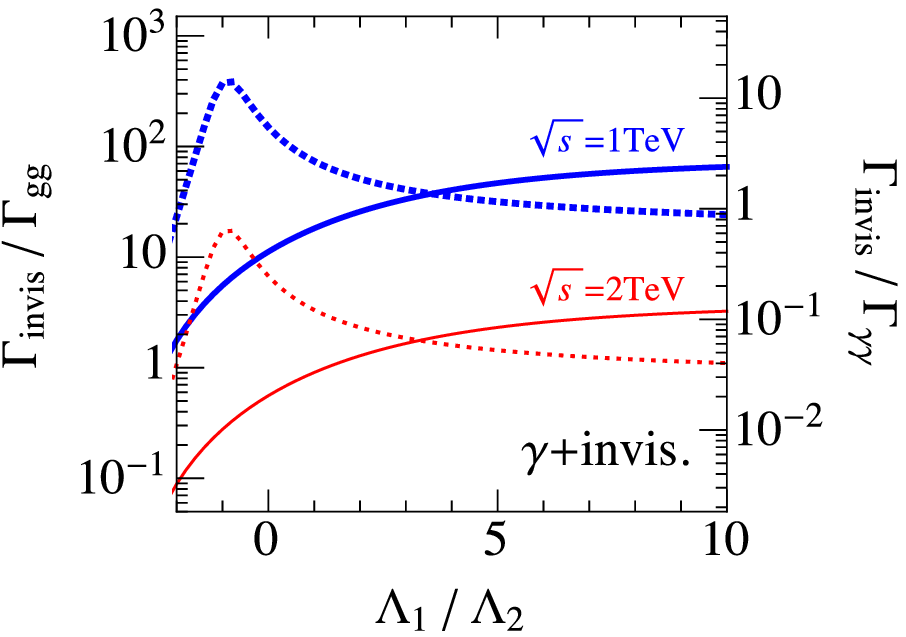}
  \caption{\small The minimal values of $\Gamma
    (\Phi\rightarrow\chi\chi)/\Gamma (\Phi\rightarrow gg)$ (left
    horizontal axis, for the case where the LHC di-photon excess is
    due to the gluon-gluon fusion) and $\Gamma
    (\Phi\rightarrow\chi\chi)/\Gamma (\Phi\rightarrow\gamma\gamma)$
    (right horizontal axis, for the case where the LHC di-photon
    excess is due to the photon-photon fusion) to realize $S_{\gamma
      \chi\chi} / \sqrt{B_{\gamma\chi\chi}} > 5$, using the production
    process of $e^+ e^- \rightarrow \Phi\gamma$.  Here, we take
    $\sqrt{s}=1\ {\rm TeV}$ (thick blue) and $2\ {\rm TeV}$ (thin
    red), and $L=1 \, {\rm ab}^{-1}$.  In addition, the helicities of
    $e^\pm$ are $(P_{e^+}, P_{e^-})=(-0.3, -0.8)$ (solid) and $(+0.3,
    -0.8)$ (dashed).}
  \label{fig:minratio_invis_a}
  \includegraphics[height=0.4\textheight]{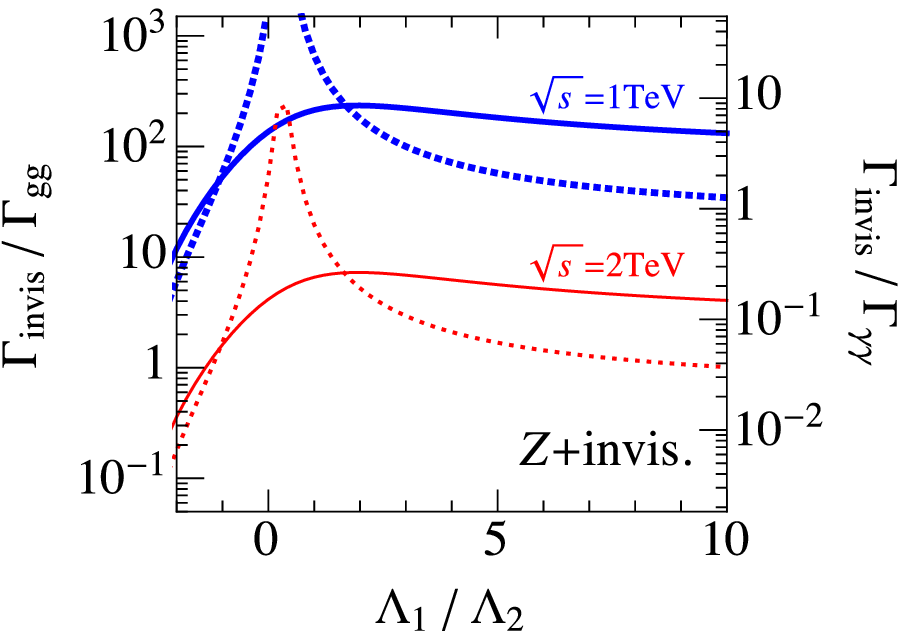}
  \caption{\small Same as Fig.\ \ref{fig:minratio_invis_a}, except for
    $e^+ e^- \rightarrow \Phi Z$.}
  \label{fig:minratio_invis_z}
\end{figure}

The detectability of the invisible decay mode is studied by
calculating the following quantity:
\begin{align}
  S_{V\chi\chi} / \sqrt{B_{V\chi\chi}} \equiv
  \frac{L \sigma (e^+ e^-\rightarrow \Phi V)
    Br (\Phi\rightarrow\chi\chi) \epsilon}
  {\sqrt{B_{V\chi\chi}}},
\end{align}
where $B_{V\chi\chi}$ is the total number of the backgrounds for the
process $e^+ e^- \rightarrow \Phi V$ followed by the invisible decay
of $\Phi$.  When the gluon-gluon fusion process dominates the LHC
di-photon signal events, $S_{V\chi\chi} / \sqrt{B_{V\chi\chi}}$ is
proportional to the ratio of $\Gamma (\Phi\rightarrow\chi\chi)/\Gamma
(\Phi\rightarrow gg)$, as can be understood from Eq.\
\eqref{sigma_gg}.  On the other hand, it is proportional $\Gamma
(\Phi\rightarrow\chi\chi)/\Gamma (\Phi\rightarrow\gamma\gamma)$ for
the case where the photon-photon fusion process is the origin of the
LHC di-photon excess (see Eq.\ \eqref{sigma_gmmgmm}).  We have
estimated the minimal values of these ratios as functions of
$\Lambda_1/\Lambda_2$ to observe the invisible decay of $\Phi$ at the
level of $S_{V\chi\chi} / \sqrt{B_{V\chi\chi}}>5$.  The results are
shown in Figs.\ \ref{fig:minratio_invis_a} and
\ref{fig:minratio_invis_z} for $e^+ e^- \rightarrow \Phi \gamma$ and
$\Phi Z$, respectively.  For the case where the di-photon excess at
the LHC originates from the gluon-gluon fusion process, for example,
the ILC with $\sqrt{s}=1\ {\rm TeV}$ and $L=1\ {\rm ab}^{-1}$ may
observe the invisible decay of $\Phi$ when
$\Gamma(\Phi\rightarrow\chi\chi)/\Gamma(\Phi\rightarrow gg)\sim
O(10)$.

\section{Conclusions and Discussion}
\label{sec:summary}
\setcounter{equation}{0}

In this paper, we have studied the prospect of investigating the
scalar boson $\Phi$, which is responsible for the di-photon excess
observed at the LHC, using the future $e^+e^-$ colliders.  We have
concentrated on the case where the LHC di-photon excess originates
from the gluon-gluon and/or photon-photon fusion processes.  We
assumed that there exists a scalar boson $\Phi$ with its mass of $\sim
750\ {\rm GeV}$, and that $\Phi$ directly couples to the SM gauge
bosons via the dimension-five operators (see Eq.\ \eqref{L_pseudo}).

We have studied the production process of $\Phi$ in association with a
SM gauge boson (i.e., $\gamma$ or $Z$) and the production via the
vector-boson fusion.  We have calculated the cross sections of these
processes.  Then we have investigated the detectability of $\Phi$ with
estimating the SM backgrounds.  Detection of the decay mode of $\Phi$
into the gluon pair seems difficult because the number of backgrounds
is large.  With the vector-boson fusion process, the decay of $\Phi$
into the electroweak gauge bosons may be detected at the $e^+e^-$
colliders if the decay width of such processes are of the same order
of $\Gamma (\Phi\rightarrow gg)$, assuming that the LHC di-photon
excess is due to the gluon-gluon fusion.  In order to observe $\Phi$,
the detection of energetic $e^\pm$ in the forward directions is
crucial to eliminate the backgrounds.  We found that the observations
of the associated production with $\gamma$ or $Z$ are more difficult
because the numbers of backgrounds are order of magnitude larger.  We
have also studied the possibility of detecting the invisible decay of
$\Phi$ using the production process in association with $\gamma$ or
$Z$.

We comment on another possibility to study $\Phi$ at the future
$e^+e^-$ facilities.  Because $\Phi$ couples directly to the photon
pair, it can be produced at the photon-photon collider which is an
important option of the future $e^+e^-$ facilities.  Indeed, in
\cite{Ito:2016zkz}, it was shown that, even if the LHC di-photon
excess originates from the gluon-gluon fusion, the $\Phi$ production
cross section at the photon-photon collider can be as large as $\sim
100\ {\rm fb}$ for $\sqrt{s}\sim 1\ {\rm TeV}$, which is much larger
than that with the $e^+e^-$ collision.  With such a large cross
section, a significant number of $\Phi$ will be available for its
detailed study.

Based on our analysis, the expected number of $\Phi$ production at the
ILC is of $O(10)$ with $\sqrt{s}=1\ {\rm TeV}$ and $L=1\ {\rm
  ab}^{-1}$, if the LHC di-photon excess is due to the gluon-gluon
fusion and also if $\Phi$ dominantly decays into the gluon pair.  If
the LHC di-photon excess is from the photon-photon fusion, the number
of $\Phi$ produced at the ILC becomes larger.  In addition, even if
the $\Phi$ production at the LHC is dominated by the gluon-gluon
fusion, the invisible decay of $\Phi$ may be observed at the ILC if
the invisible decay width is an order of magnitude larger than the
decay width into the gluon pair.  Thus, the ILC will provide
interesting possibilities to study the properties of the di-photon
resonance.

\vspace{5mm}
\noindent {\it Acknowledgment}: The authors are grateful to Y. Takaesu
for the collaboration at the early stage of this project. They also
thank K. Fujii and T. Tanabe for useful comments.  The work of T.M. is
supported by JSPS KAKENHI No.\ 26400239.

\appendix

\section*{Appendix: Cross Sections}
\setcounter{equation}{0}
\renewcommand{\theequation}{A.\arabic{equation}}

In this Appendix, we give the expressions for the cross sections of
the resonance production in association with SM gauge bosons.  For the
case with pseudo-scalar resonance, we adopt the interaction terms
given in Eq.\ \eqref{L_pseudo}.  For completeness, we also consider
the case where the new scalar resonance is a scalar boson, for which
the interaction terms are given by\footnote
{For the notational simplicity, we use the same notation for the
  suppression scales of the dimension-five operators in the
  pseudo-scalar and scalar cases.}
\begin{align}
  {\cal L}_{\rm eff} = 
  \frac{1}{\Lambda_1} \Phi
  {\cal B}_{\mu\nu} {\cal B}_{\mu\nu}
  + \frac{1}{\Lambda_2} \Phi
  {\cal W}_{\mu\nu}^a {\cal W}_{\mu\nu}^a
  + \frac{1}{\Lambda_3} \Phi
  {\cal G}_{\mu\nu}^A {\cal G}_{\mu\nu}^A.
  \label{L_scalar}
\end{align}
The following formulae can be used for both pseudo-scalar and scalar
cases.

Then, denoting the angle between the beam axis and the direction of
the $\Phi$ in the CM frame as $\theta$, the differential cross
sections of the $\Phi$ production in association with a SM gauge boson
with fully-polarized initial-state $e^\pm$ is given by
\begin{align}
  \frac{d \sigma(e^+_{R} e^-_{L} \rightarrow \Phi V)}{d \cos\theta} = 
  \frac{\beta}{32 \pi} C_V^{(L)} 
  \left[
    s^2 \beta^2 ( 1 + \cos^2 \theta ) + 8 \xi_\Phi s m_V^2
  \right],
\end{align}
with $V=\gamma$ and $Z$; $d \sigma(e^+_{L} e^-_{R} \rightarrow \Phi
V)/d \cos\theta$ can be obtained by exchanging $L\leftrightarrow R$,
while $\sigma(e^+_{L} e^-_{L} \rightarrow \Phi V)= \sigma(e^+_{R}
e^-_{R} \rightarrow \Phi V)=0$.  Here, $\xi_\Phi=0$ for the
pseudo-scalar production processes, while $\xi_\Phi=1$ for the
scalar production, $m_V=0$ and $m_Z$ for $V=\gamma$ and $Z$,
respectively, and
\begin{align}
  \beta = \frac{1}{s}
  \sqrt{s^2 - 2 (m_\Phi^2 + m_V^2) s + (m_\Phi^2 - m_V^2)^2}.
\end{align}
In addition, 
\begin{align}
  C_\gamma^{(L,R)} =
  \left[
    \frac{2e}{\Lambda_{\gamma\gamma}s}
    - \frac{g_{Ze}^{(L,R)}}{\Lambda_{\gamma Z} (s-m_Z^2)}
  \right]^2,
  \\
  C_Z^{(L,R)} =
  \left[
    \frac{e}{\Lambda_{\gamma Z}s}
    - \frac{2 g_{Ze}^{(L,R)}}{\Lambda_{Z Z} (s-m_Z^2)}
  \right]^2,
\end{align}
where 
\begin{align}
  \Lambda_{\gamma\gamma}^{-1} \equiv & \,
  \frac{g_1^2}{g_Z^2} \Lambda_2^{-1} + \frac{g_2^2}{g_Z^2} \Lambda_1^{-1},
  \\
  \Lambda_{\gamma Z}^{-1} \equiv & \,
  \frac{2 g_1 g_2}{g_Z^2} 
  (\Lambda_2^{-1} - \Lambda_1^{-1}),
  \\
  \Lambda_{ZZ}^{-1} \equiv & \,
  \frac{g_2^2}{g_Z^2} \Lambda_2^{-1} + \frac{g_1^2}{g_Z^2} \Lambda_1^{-1},
\end{align}
with $g_1$ and $g_2$ being the gauge coupling constants of $U(1)_Y$
and $SU(2)_L$, respectively, $g_Z\equiv\sqrt{g_1^2+g_2^2}$,\footnote
{In our numerical calculations, we use the gauge coupling constants at
  the renormalization scale of $\mu=m_Z$.  For more accurate
  calculations, inclusion of the effects of renormalization group
  running of the gauge coupling constants are relevant.}
and
\begin{align}
  e = \frac{g_1 g_2}{g_Z},~~~
  g_{Ze}^{(L)} = 
  \frac{g_1^2 - g_2^2}{2g_Z}, ~~~
  g_{Ze}^{(R)} = 
  \frac{g_1^2}{g_Z}.
\end{align}


\begin{thebibliography}{99}

\bibitem{Behnke:2013xla}
  T.~Behnke {\it et al.},
  arXiv:1306.6327 [physics.acc-ph];
\bibitem{Baer:2013cma}
  H.~Baer {\it et al.},
  arXiv:1306.6352 [hep-ph];
\bibitem{Adolphsen:2013jya} 
  C.~Adolphsen {\it et al.},
  arXiv:1306.6353 [physics.acc-ph];
\bibitem{Adolphsen:2013kya}
  C.~Adolphsen {\it et al.},
  arXiv:1306.6328 [physics.acc-ph];
\bibitem{Behnke:2013lya} 
  T.~Behnke {\it et al.},
  arXiv:1306.6329 [physics.ins-det].


\bibitem{Linssen:2012hp}
  L.~Linssen, A.~Miyamoto, M.~Stanitzki and H.~Weerts,
  arXiv:1202.5940 [physics.ins-det].

\bibitem{ATLAS-CONF-2015-081}
  The ATLAS Collaboration,
  ATLAS-CONF-2015-081 (2015).

\bibitem{CMS-PAS-EXO-15-004}
  The CMS Collaboration,
  CMS PAS EXO-15-004 (2015).

%
%

\bibitem{Harigaya:2015ezk}
  K.~Harigaya and Y.~Nomura,
  Phys.\ Lett.\ B {\bf 754} (2016) 151
  [arXiv:1512.04850 [hep-ph]].


\bibitem{Mambrini:2015wyu}
  Y.~Mambrini, G.~Arcadi and A.~Djouadi,
  Phys.\ Lett.\ B {\bf 755} (2016) 426
  [arXiv:1512.04913 [hep-ph]].

\bibitem{Backovic:2015fnp}
  M.~Backovic, A.~Mariotti and D.~Redigolo,
  JHEP {\bf 1603} (2016) 157
  [arXiv:1512.04917 [hep-ph]].

\bibitem{Angelescu:2015uiz}
  A.~Angelescu, A.~Djouadi and G.~Moreau,
  Phys.\ Lett.\ B {\bf 756} (2016) 126
  [arXiv:1512.04921 [hep-ph]].

\bibitem{Nakai:2015ptz}
  Y.~Nakai, R.~Sato and K.~Tobioka,
  arXiv:1512.04924 [hep-ph].

\bibitem{Knapen:2015dap}
  S.~Knapen, T.~Melia, M.~Papucci and K.~Zurek,
  arXiv:1512.04928 [hep-ph].

\bibitem{Buttazzo:2015txu}
  D.~Buttazzo, A.~Greljo and D.~Marzocca,
  Eur.\ Phys.\ J.\ C {\bf 76} (2016) no.3,  116
  [arXiv:1512.04929 [hep-ph]].

\bibitem{Pilaftsis:2015ycr}
  A.~Pilaftsis,
  Phys.\ Rev.\ D {\bf 93} (2016) no.1,  015017
  [arXiv:1512.04931 [hep-ph]].

\bibitem{Franceschini:2015kwy}
  R.~Franceschini {\it et al.},
  JHEP {\bf 1603} (2016) 144
  [arXiv:1512.04933 [hep-ph]].

\bibitem{DiChiara:2015vdm}
  S.~Di Chiara, L.~Marzola and M.~Raidal,
  arXiv:1512.04939 [hep-ph].

\bibitem{Higaki:2015jag}
  T.~Higaki, K.~S.~Jeong, N.~Kitajima and F.~Takahashi,
  Phys.\ Lett.\ B {\bf 755} (2016) 13
  [arXiv:1512.05295 [hep-ph]].



\bibitem{Ito:2016zkz}
  H.~Ito, T.~Moroi and Y.~Takaesu,
  Phys.\ Lett.\ B {\bf 756} (2016) 147
  [arXiv:1601.01144 [hep-ph]].

\bibitem{Sonmez:2016xov}
  N.~Sonmez,
  arXiv:1601.01837 [hep-ph].

\bibitem{Djouadi:2016eyy}
  A.~Djouadi, J.~Ellis, R.~Godbole and J.~Quevillon,
  JHEP {\bf 1603} (2016) 205
  [arXiv:1601.03696 [hep-ph]].

\bibitem{He:2016olo}
  M.~He, X.~G.~He and Y.~Tang,
  arXiv:1603.00287 [hep-ph].


\bibitem{Csaki:2015vek}
  C.~Csaki, J.~Hubisz and J.~Terning,
  Phys.\ Rev.\ D {\bf 93} (2016) 035002
  [arXiv:1512.05776 [hep-ph]].

\bibitem{Csaki:2016raa}
  C.~Csaki, J.~Hubisz, S.~Lombardo and J.~Terning,
  arXiv:1601.00638 [hep-ph].

\bibitem{Aad:2014fha}
  G.~Aad {\it et al.} [ATLAS Collaboration],
  Phys.\ Lett.\ B {\bf 738} (2014) 428
  [arXiv:1407.8150 [hep-ex]].

\bibitem{Budnev:1974de}
  V.~M.~Budnev, I.~F.~Ginzburg, G.~V.~Meledin and V.~G.~Serbo,
  Phys.\ Rept.\  {\bf 15} (1975) 181.

\bibitem{Gounaris:1999gh} 
  G.~J.~Gounaris, P.~I.~Porfyriadis and F.~M.~Renard,
  Eur.\ Phys.\ J.\ C {\bf 9}  (1999) 673
  [hep-ph/9902230].
  
\bibitem{Gounaris:1999ux} 
  G.~J.~Gounaris, J.~Layssac, P.~I.~Porfyriadis and F.~M.~Renard,
  Eur.\ Phys.\ J.\ C {\bf 10}  (1999) 499
  [hep-ph/9904450].

\bibitem{Gounaris:1999hb}
  G.~J.~Gounaris, J.~Layssac, P.~I.~Porfyriadis and F.~M.~Renard,
  Eur.\ Phys.\ J.\ C {\bf 13} (2000) 79
  [hep-ph/9909243].

\bibitem{Alwall:2014hca} 
  J.~Alwall {\it et al.},
  JHEP {\bf 1407}, 079 (2014)
  [arXiv:1405.0301 [hep-ph]].

\bibitem{Conte:2012fm} 
  E.~Conte, B.~Fuks and G.~Serret,
  Comput.\ Phys.\ Commun.\  {\bf 184}, 222 (2013)
  [arXiv:1206.1599 [hep-ph]].










\end{thebibliography}
\end{document}